\documentclass[prx, aps, twocolumn, amsmath, amssymb, floatfix]{revtex4-2}

\usepackage{graphicx}
\usepackage{color}
\usepackage[normalem]{ulem}
\usepackage{epstopdf}
\usepackage{lmodern}
\usepackage{microtype}
\usepackage[utf8]{inputenc}
\usepackage{lipsum}
\usepackage[colorlinks=true,citecolor=blue,linkcolor=blue,urlcolor=blue]{hyperref}
\usepackage{siunitx}

\usepackage{tikz}
\usepackage{graphicx}
\usepackage{braket}

\begin{document}
\title{
Optimal squeezing for high-precision atom interferometers
}

\author{P. Feldmann$^{1,2,3}$}
\author{F. Anders$^4$}
\author{A. Idel$^{4}$}
\author{C. Schubert$^{4,5}$}
\author{D. Schlippert$^{4}$}
\author{L. Santos$^2$}
\author{E.\,M. Rasel$^{4}$}
\author{C. Klempt$^{4,5}$}
\email[Author to whom correspondence should be addressed: ]{carsten.klempt@dlr.de}

\affiliation{$^1$Leibniz Universit\"at Hannover, Institut f\"ur Theoretische Physik, Appelstra\ss{}e~2, D-30167~Hannover, Germany\\ 
$^2$Stewart Blusson Quantum Matter Institute, The University of British Columbia, 2355 East Mall, Vancouver BC V6T 1Z4, Canada\\
$^3$Department of Physics \& Astronomy, The University of British Columbia, 6224 Agricultural Road, Vancouver BC V6T 1Z1, Canada\\
$^4$Leibniz Universit\"at Hannover, Institut für Quantenoptik, Welfengarten 1, D-30167 Hannover, Germany  \\
$^5$Deutsches Zentrum für Luft- und Raumfahrt (DLR), Institut für Satellitengeod\"asie und Inertialsensorik, Callinstr. 30b, D-30167 Hannover, Germany}

\date{\today}

\begin{abstract}
We show that squeezing is a crucial resource for interferometers based on the spatial separation of ultra-cold interacting matter.
Atomic interactions lead to a general limitation for the precision of these atom interferometers, which can neither be surpassed by larger atom numbers nor by conventional phase or number squeezing.
However, tailored squeezed states allow to overcome this sensitivity bound by anticipating the major detrimental effect that arises from the interactions.
We envisage applications in future high-precision differential matter-wave interferometers, in particular gradiometers, e.g., for gravitational-wave detection.
\end{abstract}

\maketitle

\section{Introduction}
Interferometers based on ultra-cold interacting atoms are at the very heart of the second quantum revolution.
A usual assumption is that their sensitivity scales with the number of employed atoms.
We show that atomic interactions affect and eventually limit the resolution of these interferometers due to the quantum noise inherently linked to splitting atomic ensembles.
This effect results in a maximal useful number of particles in the atomic ensemble a corresponding sensitivity limit.
To overcome this constraint, we propose to utilize tailored squeezed states.

\begin{figure}[t]
    \centering
    \includegraphics[width=0.8\columnwidth]{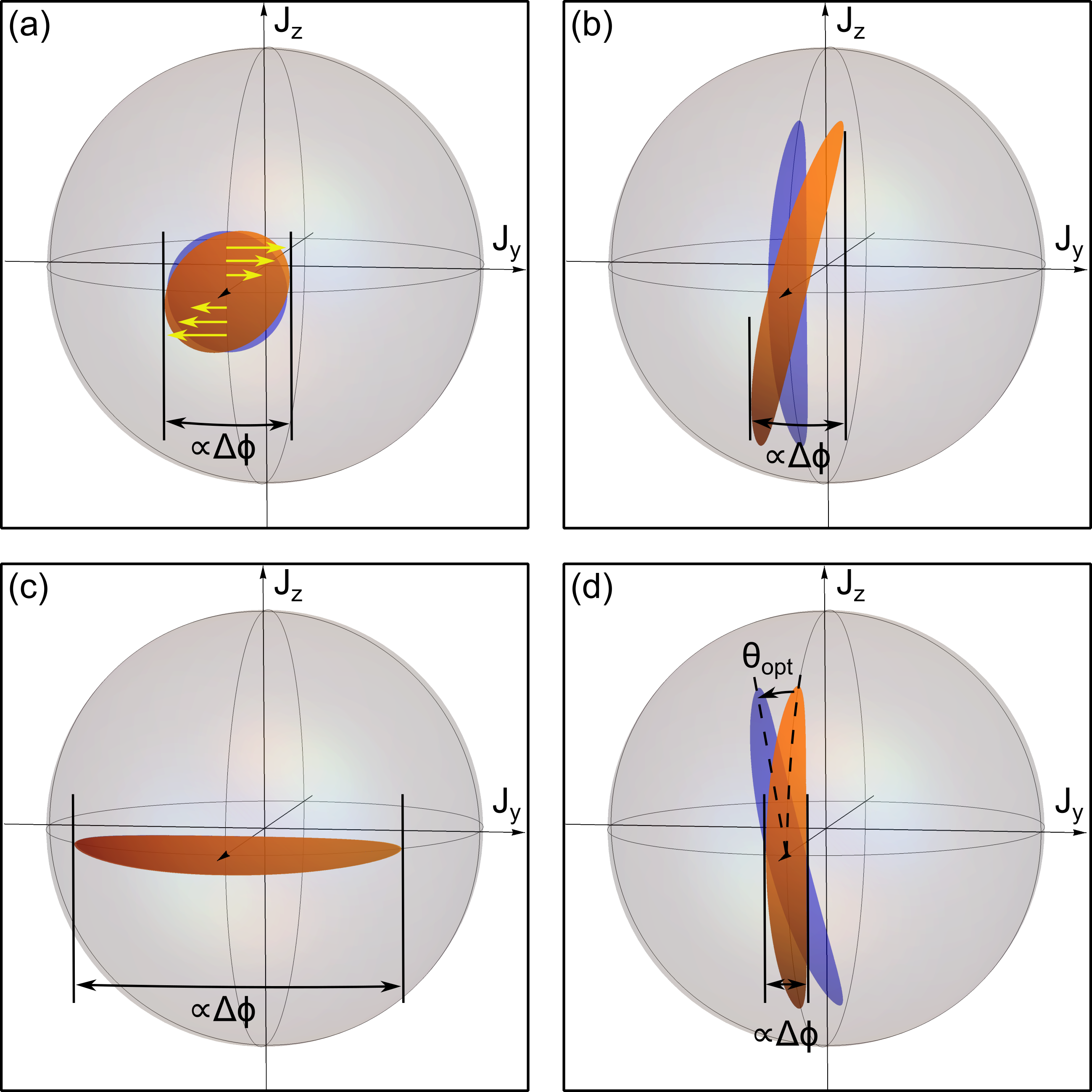}
    
    	\caption{
    \label{fig1}
    Different states in an atom interferometer represented on the many-particle Bloch sphere.
    The density effect (yellow arrows) alters the input states (blue ellipses) into the states that actually sense the interferometric phase shift (orange ellipses).
    The optimal phase uncertainty is proportional to the resulting width in $J_y$ direction. 
    (a) A coherent state evolves into a mildly squeezed and improperly oriented state, which increases the noise in the phase readout.
    (b) A phase-squeezed state loses part of its quantum-enhanced phase sensitivity. The squeezing orientation is effectively rotated and thereby the anti-squeezed uncertainty couples into the phase measurement.
    (c) The number-squeezed state barely changes but suffers from initial anti-squeezing.
    (d) Optimal-orientation squeezing takes into account the effective rotation by the density effect such that the state becomes optimally orientated.
    For underlying physical parameters see Fig.~\ref{fig3}.
    To improve the visualization, the size of the states in proportion to the sphere is exaggerated.
    }
\end{figure}

Atom interferometers, in particular light-pulse interferometers, enable a wide range of applications such as inertial sensing, measurements of the photon-recoil and the gravitational constant, gravimetry, gravity gradiometry~\cite{Asenbaum2017,Chiow2016,Biedermann2015,McGuirk2002,McGuirk2000}, and tests of general relativity~\cite{werner_atom_2023,overstreet_observation_2022,asenbaum_atom-interferometric_2020}. They were moreover proposed for the detection of ultralight dark matter and gravitational waves (GWs)~\cite{abend_terrestrial_2023,Abe_MAGIS_2021,Badurina_AION_2020,Canuel2020,Canuel2018,Hogan2016,Hogan2011,Dimopoulos2008}.

Unprecedented resolutions are anticipated for the two latter applications as they are based on differential measurements, which allows for an efficient suppression of common-mode noise.
The measurement resolution is fundamentally limited by the projection noise of the employed quantum state, which is commonly improved by increasing the atom number $N$.
In addition to a large $N$, the ensembles in such interferometers also need to be extremely cold and remain compact in order to reduce the impact of environmental effects such as Earth's rotation $\Omega$ or gravity gradients $\gamma$. Otherwise,
the uncertainties in the external degrees of freedom of the atoms, i.e., in position, $\Delta r$, and velocity, $\Delta v$, would compromise the sensitivity unless the environmental effects are compensated to a challenging degree.
Increasing the atom number while keeping the wave packets compact, however, makes the influence of atomic interactions crucial for precision measurements.

Even at negligible technical noise, the density-dependent interactions in a spatially split ensemble unavoidably lead to an increased quantum projection noise that deteriorates the phase readout (see Fig.~\ref{fig1}a).
For large atom numbers and a given extension of wave packets, the trade-off between a decreasing projection noise and an increasing density-dependent phase noise leads to a maximal useful ensemble size $N$ and a limitation on the achievable resolution, which we will denote as density quantum limit (DQL).

Complementary to increasing the atom number, entanglement, e.g. squeezed states, can reduce the relevant projection noise.
However, conventional squeezed states are even more prone to the density effect, as the resulting anti-squeezing is coupled into the measurement.
If a squeezed state is prepared in the usually optimal phase-squeezing direction (light blue ellipse in Fig.~\ref{fig1}b), the atomic interaction leads to a twist, such that the anti-squeezed number-noise is rotated into the phase readout.
In this work, we show that the effect of interactions can be anticipated and hence canceled by squeezing the ensemble's quadratures in an appropriate orientation between the ones corresponding to phase and number squeezing (Fig.~\ref{fig1}d).
This method of optimal-orientation squeezing (OOS) can push the sensitivity of atom interferometers beyond the DQL.

While our method is generally applicable, our experimental proposal considers light-pulse atom interferometers based on rubidium, which are commonly employed for many applications and offer various squeezing concepts.
We follow the approach to generate entanglement in spin space and to subsequently transfer the entangled modes to momentum states, as it relies on well-established methods in atom interferometry like Raman processes.
Such a generation of entanglement in momentum space was demonstrated for Bose-Einstein condensates~\cite{Anders2021} and for thermal clouds~\cite{Greve2022,Malia2022}, where it was also implemented in an atom interferometer.
Our method is also applicable when the entanglement is directly generated between momentum modes by, e.\,g., delta-kick squeezing~\cite{Corgier2021}.

The article is organized as follows:
In Section~\ref{sec:optimal_squeezing} we describe how we model the density effect in atom interferometers.
The sensitivity bound of the DQL and the beneficial effect of OOS are discussed in Section~\ref{sec:DQL}.
Evaluating the relevant noise sources, we show in Section~\ref{sec:quantum limits} that near-term differential inertially sensitive atom interferometers will be limited by the DQL.
A specific realization of a squeezing-enhanced gravity gradiometer or gravitational-wave detector is proposed in Section~\ref{sec:squeezing_proposal}, before we close with a conclusion and outlook in Section~\ref{sec:conclusion}.

\section{Density effect in atom interferometers}
\label{sec:optimal_squeezing}

We analyze the impact of interactions on squeezed probes of two-mode atom interferometers~\cite{Pezze2018}.
As an atomic source we assume a Bose-Einstein condensate (BEC), benefiting from low expansion velocities and excellent mode control which are required for large-scale atom interferometers~\cite{Hensel2021epjd,Schubert2019,Karcher2018NJP,Hartwig2015,Dickerson2013,Szigeti2012}.
An atom in two modes $a$ and $b$ can be represented by a spin of $1/2$.
Then the quantum state of $N$ indistinguishable bosonic atoms has a total spin $\vec{J}$ of length $N/2$ and can be visualized on the many-particle Bloch sphere of radius $N/2$ (Fig.~\ref{fig1}).
Each axis of the Bloch sphere corresponds to a component of the collective spin with $J_x=(a^\dagger b + b^\dagger a)/2$, $J_y=(a^\dagger b -  b^\dagger a)/(2i)$, $J_z=(a^\dagger a - b^\dagger b)/2$.

\begin{figure*}[ht!]
	\centering
	\includegraphics[width=1.\textwidth]{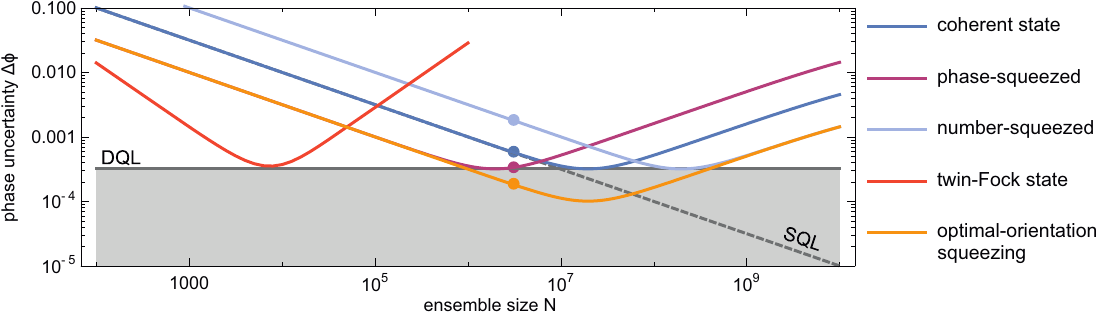}
		\caption{
		\label{fig3}	
		Phase uncertainty including the density effect as a function of the atom number.
		The performance of the coherent state (blue) as well as the phase-squeezed (purple) and number-squeezed (light blue) states scales with $\xi/\sqrt{N}$ for relatively small $N$ and with at most $\sqrt{N}/\xi$ for large ensembles.
		The SQL (dashed gray) is at $1/\sqrt{N}$.
		None of the coherent state (blue), the conventionally squeezed states (purple, light blue), or the highly entangled twin-Fock state (red) achieves phase uncertainties below the DQL (gray area).
		However, for OOS (orange), i.e., squeezing that anticipates the density effect, the sensitivity surpasses the DQL.
		Here we use the gradiometer parameters from Fig.~\ref{fig2}c and assume $\SI{10}{dB}$ of squeezing. The colored dots correspond to the states visualized in Fig.~\ref{fig1}.
		}
\end{figure*}

Similarly to the classical Ramsey scheme, atoms are prepared at the north pole of the Bloch sphere and rotated around $J_y$ onto the equator by a symmetric beam splitter. 
Here, the atoms are sensitive to a phase shift $\phi$ around $J_z$, which is induced by the quantity of interest (such as gravity) during  a variable evolution time.
After a final beam splitter (around $J_x$), the result can be read out by a measurement along the $J_z$ axis.
For small $\phi$, the measurement uncertainty can reach down to 
\begin{equation}
    \label{eq:sensitivity}
    \Delta^2\phi =\Delta^2J_y/\langle J_x\rangle^2,
\end{equation}
where variance and expectation value are evaluated with respect to the state $|\psi\rangle$ pointing along the $J_x$-axis and obtained after the first beam splitter (blue ellipses in Fig.~\ref{fig1}).
For the classical Ramsey scheme with all atoms initially in mode $a$, the resolution is bounded by the standard quantum limit (SQL) $\Delta^2\phi_{\mathrm{SQL}} = 1/N$.
For a squeezed initial state, Eq.~\eqref{eq:sensitivity} evaluates to $\Delta^2\phi =\xi^2/N < 1/N$.
The SQL can thus be surpassed by a constant factor by, e.g., initially populating mode $b$ with a squeezed vacuum.

While the SQL improves with increasing $N$, density effects grow larger and commonly decrease the overall phase sensitivity.
Interactions among $N$ bosons uniformly occupying a volume $V$ lead to a mean-field shift of the ground-state energy $\propto N^2/V$.
When the two modes $a$ and $b$ get spatially separated after the first beam splitter, atoms in each mode experience their own density-dependent shift.
Since even an ideal splitting process induces quantum fluctuations of the relative population of the modes, the relative phase shift, zero on average, becomes noisy.
The larger the population imbalance $N_a-N_b=2 J_z$ of the two modes, the larger is the relative phase shift, i.e. the effect resembles a $J_z$-dependent rotation around $J_z$ on the Bloch sphere.
Indeed, the density effect can be modeled by the one-axis twisting Hamiltonian $H=\hbar \chi J_z^2$~\cite{Kitagawa1993}.
The overall strength of the effect is captured by $\mu=2\int_0^{2T}\!\operatorname{d}\!t\,\chi(t)$, where $2T$ is the interrogation time of the atom interferometer and
 $\hbar\chi=U/V$ with $U=4\pi\hbar^2a_s/m$, $a_s$ the scattering length and $m$ the mass of the atoms~\cite{Pitaevskii2003}.
Importantly, $H$ commutes with the phase imprinting.
Therefore, the sensitivity in the presence of such a density effect is given by Eq.~\eqref{eq:sensitivity} with respect to $e^{-i\mu J_z^2/2}|\psi\rangle$ (orange ellipses in Fig.~\ref{fig1}).

\section{Surpassing the density quantum limit by optimal-orientation squeezing}
\label{sec:DQL}

Figure~\ref{fig1} visualizes the density effect for various input states on the Bloch sphere.
In (a), the density effect slightly twists a symmetric coherent state into a tilted ellipse, resulting in an increased measurement uncertainty $\Delta \phi$ along $J_y$.
The initial $J_y$-uncertainty can be minimized by a phase-squeezed state (b), but the density effect couples the anti-squeezed uncertainty along $J_z$ into the measurement direction $J_y$. 
On the other hand, the number-squeezed state in (c) is only barely affected by the density effect but suffers from the initially anti-squeezed phase uncertainty. Squeezing remains useful even if the density effect is not negligible (d). However, to account for the twisting, the squeezing direction has to be appropriately chosen between phase squeezing (squeezing angle $\theta=0$) and number squeezing ($\theta=\pi$). 

In Figure~\ref{fig3} we show how OOS allows to surpass the DQL for a realistic gradiometer configuration detailed below. It shows the phase-estimation uncertainty as a function of atom number $N$ for various initial states. We approximate the coherent and squeezed probe states by one-mode quadrature-squeezed Gaussian states; the twin-Fock state is treated exactly. As a sanity check, we have compared the approximate phase-estimation uncertainty for a coherent state with the exact result and confirmed that the difference is negligible. To describe the spatial expansion, we use Gaussian wave functions as a variational ansatz (see Ref.~\cite{PerezGarcia1996} and Appendix~\ref{sec:gaussvariation}) and approximate the atom number with $N/2$ per wave packet.

For a coherent probe state (blue line), the uncertainty drops with $1/\sqrt{N}$ (dashed gray line) until the density effect dominates, leading to an increase with $\sqrt{N}$ for large ensembles.
The minimal uncertainty defines the DQL, which is here reached at a total atom number of $N \approx \SI{2e7}{}$. Phase-squeezed input states (purple line) enable entanglement-enhanced measurements (beyond the SQL) for relatively small ensembles, but the sensitivity deteriorates for large atom numbers because the density effect couples the anti-squeezed number noise into the measurement.
The optimal resolution is obtained at a smaller $N$ compared to the coherent state but does not surpass the DQL.
Input states with squeezed number noise (light blue line) do not allow for a sensitivity beyond the SQL. However, they reach the DQL at larger atom numbers and therefore outperform the coherent state and the phase-squeezed state in the regime where the density effect dominates.
The DQL also poses a barrier for some much more entangled states, as we demonstrate taking the twin-Fock state as an example (red line).
The twin-Fock state offers a phase uncertainty scaling as the Heisenberg limit $\Delta^2 \phi \propto 1/N^2$~\cite{Bouyer1997,Lucke2011} before it suffers from a strong density effect.

The maximal sensitivity at the DQL can be approximated as
\begin{equation}
    \label{eq:DQL}
    \Delta^2\phi_{\mathrm{DQL}}\approx\frac{3U\bar{m}y}{4\pi\hbar\Delta r }(1-2cy)
\end{equation}
with $c=\bar{m}\Delta r\Delta v$ and $y = T/[4cT + 2\bar{m}(\Delta r)^2]$. Equation~\eqref{eq:DQL} relies on the one-mode and the single-spatial-mode approximations. Moreover, we have assumed that the interactions negligibly contribute to the expansion of wave packets and that $y\ll 1$, see Appendix~\ref{sec:app} for a detailed derivation. The DQL represents a general sensitivity limit which may serve as a reference for the multitude of high-precision atom interferometry projects world-wide, both for actual measurements as well as planned terrestrial or space-borne missions.

Remarkably, using OOS (orange line in Fig.~\ref{fig3}) is a simple way to surpass the DQL by the full amount of inital squeezing. 
In one-mode approximation, i.\,e., for large atom numbers and moderate squeezing, the squeezing angle of OOS is given by
\begin{equation}
    \label{eq:optimalsqz}
    \tan\theta_{\mathrm{opt}}=\frac{4\mu N}{4-\mu^2 N^2}
\end{equation}
with $(4-\mu^2N^2) \cos \theta_{\mathrm{opt}} > 0$. The optimal resolution is reached at the same atom number $N$ at which a coherent probe would hit the DQL. 

When employing OOS, fluctuations of the total atom number cannot be fully removed by normalization, as they lead to fluctuations of the optimal orientation. The requirements for the number stability are weak for low squeezing, but become more stringent for an increased squeezing strength.

In Section~\ref{sec:optimal_squeezing}, we modeled the density effect by one-axis twisting assuming wave packets with uniform densities.
For realistic, i.\,e., inhomogeneous clouds, we expect the effect to be qualitatively similar but stronger, making OOS essential already at lower sensitivities.
In the following, we will evaluate how the presented sensitivity limitation by the DQL can be surpassed by OOS in the context of realistic high-precision atom interferometers.

\section{Sensitivity limits for inertial quantum sensing}\label{sec:quantum limits}
\begin{figure*}[htb!]
	\centering
	\includegraphics[width=.33\textwidth]{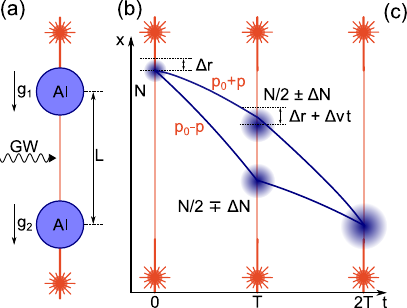}
	\scriptsize
	\begin{tabular}[b]{lccc} \\ \hline 
   \textbf{} & \textbf{Variable}                    & \textbf{Gradiometer} & \hspace{0.3em}\textbf{GW detector} \\ \hline
   Targeted sensitivity to &&&\\$\gamma$ (at 1\,s), $h$ (at peak sensitivity)  & & $6\cdot10^{-15}\,s^{-2}$ & $2\cdot10^{-21}$\\\hline
   \vphantom{\large{A}}Atom number (at 1\,s)  & $N$                       & $2\cdot 10^7$ & $6\cdot 10^{11}$ \\
   Initial radius (mm) & $\Delta r$         & 0.3 & 8 \\  
   Expansion rate ($\mu$m/s) & $\Delta v$   & 100 & 100 \\  
   Wavenumber ($2\pi/(780\,\mathrm{nm})$) & $k$ & 1000 & 2000 \\
   Free-fall time (s) &  $2T$                & 2.4 & 0.6 \\
   Baseline (m) & $L$                       & 1 & $16.3\cdot10^3$ \\ \hline
   \vphantom{\large{A}}Targeted phase sensitivity & $\Delta \phi$ & $5\cdot 10^{-5}$ & $7\cdot 10^{-7}$ \\\hline
   \vphantom{\large{A}}Standard quantum limit & $\Delta\phi_{\mathrm{SQL}}=1/\sqrt{N}$                & $2\cdot10^{-4}$ & $1\cdot10^{-6}$ \\
   Gravity gradient \& position~\cite{Loriani2020PRD,Hogan2008,Bongs2006}\hspace{0.6em} & $\Delta\phi_{\gamma,r}=k T^2 c_{\gamma}\gamma \Delta r/\sqrt{N}$ & $2\cdot10^{-5}$ & $5\cdot10^{-7}$ \\
   Gravity gradient \& velocity~\cite{Loriani2020PRD,Hogan2008,Bongs2006} & $\Delta\phi_{\gamma,v}=k T^3 c_{\gamma}\gamma \Delta v/\sqrt{N}$ & $1\cdot10^{-5}$ & $2\cdot10^{-9}$ \\
   Rotation \& velocity~\cite{Loriani2020PRD,Hogan2008,Bongs2006} & $\Delta\phi_{\Omega}=2 k T^2 c_\Omega\Omega \Delta v/\sqrt{N}$ & $4\cdot10^{-5}$ & $3\cdot10^{-5}$ \\
   Density quantum limit [Eq.~\eqref{eq:DQL}] & $\Delta \phi_{\mathrm{DQL}}$ & $3\cdot 10^{-4}$ & $2\cdot10^{-6}$
   \\ \hline
    \end{tabular}
	\caption{
        \label{fig2}
        (a)~Geometry of two atom interferometers (AIs) a distance L apart. Both AIs share the same light pulses to measure either the difference of the gravitational accelerations $g_{1,2}$ or space-time deformation due to a gravitational wave (GW).
        (b)~Space-time diagram of a Mach-Zehnder-like AI exploiting a BEC as input state with an uncertainty in position $\Delta r$ and velocity $\Delta v$.
        The first light pulse (orange lasers) generates a coherent superposition of two momentum states that spatially separate.
        (c)~Noise contributions for a gradiometer and a GW detector without quantum enhancement.
        We assume two Mach-Zehnder-like interferometers operating with $^{87}$Rb BECs in the hyperfine ground state $F=1$, resembling a symmetric Ramsey-Bord{\'e} scheme with negligible pulse separation time $T'$ between the two central pulses (cf. Fig.~\ref{fig4}), and compare various kinds of single-AI phase noise $\Delta\phi$ with the desired sensitivities to gravity gradients (Allan deviation) and strain at 1\,s~\cite{Hohensee2011,Dimopoulos2008}.
        Our parameters are gravitational acceleration $g=\SI[per-mode=symbol]{-9.78}{\metre\per\second\squared}$ \cite{Groten2000}, its gradient $\gamma\approx -2g/R$, Earth's radius $R=\SI{6378}{\kilo\metre}$ \cite{Groten2000}, Earth's angular velocity $\Omega=\SI[per-mode=symbol]{7.29e-5}{\radian\per\second}$ \cite{Groten2000}, mean scattering length $a=(2a_2+a_0)/3$ with $a_0=\SI{101.8}{a_B}$ and $a_2=\SI{100.4}{a_B}$ \cite{Kawaguchi2012}, and atomic mass $m=\si{87}{u}$. 
        We assume compensation of rotation~\cite{Hauth2013,Lan2012} and gravity gradient~\cite{Overstreet2018,Roura2017}, expressed by the suppression factors $c_{\gamma}=0.01$ (both cases) and $c_{\Omega}=0.001$ (gradiometer only).
        Our choices for $N$, $k$, $T$, and $L$ loosely resemble Refs.~\cite{Dimopoulos2007,Chaibi2016,Dimopoulos2008}. 
        The assumptions for $\Delta r$ and $\Delta v$ are constrained by the coupling to rotations~\cite{Loriani2020PRD,Hogan2008,Bongs2006}, gravity gradients~\cite{Loriani2020PRD,Hogan2008,Bongs2006}, interactions~\cite{Debs2011}, and beam-splitting efficiency~\cite{Gebbe2021NatComm,Szigeti2012}.
        Note that in the GW detector case $\Delta\phi_{\Omega}$ exceeds the DQL.
        This could be alleviated in a 4-pulse or 5-pulse geometry~\cite{Hogan2011,Schubert2019,Canuel2020}.
        }
\end{figure*}

Near-term differential atom interferometers (Fig.~\ref{fig2}a) will be able to suppress technical noise below the DQL. 
The table in Fig.~\ref{fig2}c presents the leading noise terms for typical parameters of two exemplary interferometric set-ups assuming today's state-of-the-art as well as a hypothetical very brilliant ultra-cold atom source, respectively.
A maximal wave-packet radius has to be chosen as a central parameter for the instrument design.
While a large radius suppresses density effects, it must remain small enough to suppress the impact of rotations and gravity gradients.
In both set-ups, OOS allows to surpass the density limit and to maximally benefit from the ultra-cold ensembles.

In detail, we consider two identical Mach-Zehnder-like atom interferometers  (Fig.~\ref{fig2}b) that are manipulated by common light pulses.
Each pulse transfers even multiples $k=2nk_{\mathrm{ph}}$ of photon momentum $k_{\mathrm{ph}}$ such that the atomic wave packets gain a momentum $p=\pm \hbar k$, respectively.
Atoms are coherently split, redirected, and, after a total interrogation time $2T$, recombined. 
The sharing of light pulses suppresses common noise in the differential signal.
Each interferometer is sensitive to inertial forces such as the local gravitational acceleration causing a phase shift~\cite{Hogan2008,Bongs2006}
\begin{equation}
	\label{eq:phaseGradio}
    \phi_{g}= k g T^2.
\end{equation}
Hence, the gradient $\gamma \approx (g_1-g_2)/L$ is accessible via the difference of the gravity values $g_{1}$ and $g_{2}$ that are measured by two interferometers a distance $L$ apart.

Moreover, the differential phase depends on phase shifts experienced by the light traveling over the distance $L$ and, consequently, is also modified by GWs.
In the vertical arrangement~\cite{Hogan2016,Hogan2011} suggested by Figs.~\ref{fig2}a and b, a gravitational wave of frequency $\Omega_{\text{GW}}$ and amplitude $h$ acting on the light traveling along $L$ induces a differential phase of~\cite{Hohensee2011,Dimopoulos2008}
\begin{equation}
	\label{eq:phaseGW}
    \phi_{\text{GW}}= 2 h k L \sin^2 \left( \Omega_{\text{GW}} T/2 \right).
\end{equation}
In comparison to gradiometry, the inherently weak signal of GWs  requires the scale factor $kL$ as well as---due to the SQL---the atom number $N$ to be increased significantly.
To implement very large $L$, the vertical setup can be rearranged to operate as a horizontal antenna~\cite{Canuel2018}.

For both applications, the table in Fig.~\ref{fig2}c summarizes the expected magnitude of the phase signal as well as exemplary parameters to achieve the required sensitivities using a $^{87}$Rb BEC with hyperfine spin $F=1$.
We choose the two set-ups to most prominently differ in the atom number $N$, the wave-packet extension $\Delta r$, and the distance $L$, loosely based on Refs.~\cite{Dimopoulos2007,Chaibi2016,Dimopoulos2008}. In particular, the assumed free-fall time of the GW detector bridges the frequency gap between current ground-based and space-based instruments. For the stated values of $\Delta r$, $\Delta v$ and $T$, the chosen values of $N$ are optimal in terms of the DQL.

Various noise sources can hamper precision measurements.
Here, we focus on terms depending on atom number, ensemble size, and residual expansion rate, and only briefly comment on other contributions. 
The differential measurement to a large extent rejects common-mode noise, e.g., originating from microseisms, which are otherwise typically the dominant noise source in inertially sensitive atom interferometers~\cite{McGuirk2002}.
In addition to the gravity-gradient and GW signals and next to atomic and magnetic
interactions, the arrangement is also sensitive to a variety of other effects, most prominently gravity gradients 
and rotations (Sagnac effect) that couple to the initial position and velocity distribution of the ensemble~\cite{Hogan2008,Dimopoulos2008}.
Light-pulse interferometry offers cancellation schemes for the latter two effects. 
Examples include the five-pulse geometry introduced in Ref.~\cite{Hogan2011} to suppress the Sagnac effect as well as other compensation schemes for rotations~\cite{Hauth2013,Lan2012} and gravity gradients~\cite{Overstreet2018,Roura2017}.
Imperfect cancellation, e.g., when limited knowledge constraints adaptation, restricts the atom number as well as the initial volume and expansion rate of the atomic wave packet, as indicated in Fig.~\ref{fig2}c. 

We assume that gravity gradients are compensated to $\SI{1}{\percent}$ (both cases in Fig.~\ref{fig2}c) and Earth's rotation to $\SI{0.1}{\percent}$ (gradiometer only: the long baseline of the GW detector prohibits the required dynamic range for beam steering).
Even at this level of compensation, the volumes and expansion rates required for the target sensitivities
can hardly be achieved with thermal or laser cooled ensembles.
They are, however, accessible with BECs, which offer excellent control of their spatial mode and, combined with delta-kick collimation~\cite{Ammann1997}, admit ultra-low kinetic expansion energies~\cite{Deppner2021}.
The latter are also mandatory for near-unity efficiency when transferring momentum by light-pulses~\cite{Gebbe2021NatComm,Szigeti2012}.
The resulting phase noise is summarized in Fig.~\ref{fig2}c for $N$ atoms and
position and velocity uncertainty ${\Delta}r/\sqrt{N}$ and ${\Delta}v/\sqrt{N}$, respectively~\cite{Debs2011,Hogan2008,Bongs2006,Borde1989}. 

Confronting all noise terms, the table in Fig.~\ref{fig2}c reveals that, for the suggested parameters, the sensitivity of the gradiometer is limited by the DQL. The GW detector could reach the DQL if supplemented by one order of magnitude of rotation compensation (e.g. by a 4-pulse or 5-pulse scheme~\cite{Hogan2011,Schubert2019,Canuel2020}).
Lowering the density effect by increasing the extension of the wave packets is hindered by
the uncertainties due to the Earth's gravity gradient and rotation.
Beyond balancing the various effects, the performance can only be improved by reducing the impact of interactions, for example by generating OOS as we propose in the following section.

\section{Implementing optimal-orientation squeezing in an atom interferometer}
\label{sec:squeezing_proposal}
From today’s perspective, the most promising approach to entangling different atomic momentum states is based on a protocol which first establishes entanglement in spin space and then selectively alters the momentum of the spin states by a Raman coupling~\cite{Anders2021}.
The momentum states obtained in this way intrinsically match the states traversing typical atom interferometers and can be further manipulated by Raman or Bragg processes~\cite{Mueller2009a,McDonald2014EPL,McGuirk2000,Plotkin2018PRL,Berg2015,Chiow2011,Gebbe2021NatComm,beguin_atom_2023,kirsten-siems_large-momentum-transfer_2023}.

\begin{figure}[ht!]
    \centering
	 \includegraphics[width=\columnwidth]{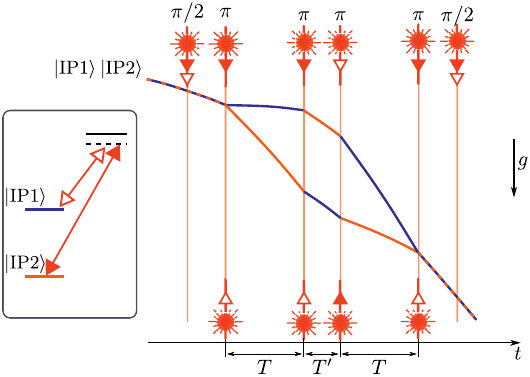}
		\caption{Single atom interferometer utilizing a BEC in an optimal-oriented squeezed state.
		Here $\ket{\text{IP}1}$ is massively populated and $\ket{\text{IP}2}$ is in a squeezed vacuum state aligned such that the sensitivity can surpass the DQL.
		The inset shows the Raman coupling of the two hyperfine levels $(F=1, m_F=0)$ (orange) and $(F=2, m_F=0)$ (purple) in Rubidium via a higher electronic state.
		The first co-propagating Raman $\pi/2$ pulse opens the interferometer and couples the two modes (rotating the state from the north pole to the equator of the Bloch sphere).
		A successive counter-propagating Raman $\pi$ pulse initiates the spatial splitting of the two modes to create an interferometer enclosing a space-time area.
		The two entangled clouds follow spatially separated paths until
		a combination of three counter-propagating Raman $\pi$ pulses reflects the trajectories and ensures spatial overlap after an interrogation time of $2T+T'$.
		Note that for reflection, the direction of the frequency components (filled and open triangle) has to be inverted for one of the pulses.
		A final co-propagating $\pi/2$ pulse turns the relative phase shift into a population imbalance of the two modes.
		For a short pulse separation time $T'$, the scheme resembles the 3-pulse geometry analyzed in Fig.~\ref{fig2}.
		By adding further sets of $\pi$ pulses separated by a short time $T'$, the geometry can be extended to resemble 4-pulse or 5-pulse schemes and exploit their suppression of spurious phase terms, e.g., due to rotations (see Fig.~\ref{fig2}c)~\cite{Hogan2011,Schubert2019,Canuel2020}.
		\label{fig4}}
\end{figure}

The following procedure prepares an optimally squeezed probe state from a $^{87}$Rb BEC in the hyperfine state $\ket{F, m_f}=\ket{1,0}$.
Spin-changing collisions convert part of the BEC to a two-mode squeezed state in the levels $\ket{1,\pm 1}$~\cite{Barnett1990,Peise2015a}.
The remaining population in $\ket{1,0}$ is transferred to $\ket{2,0}$ and forms the interferometric input state $\ket{\text{IP}1}$.

The second input state $\ket{\text{IP}2}$ is generated from the two-mode squeezed vacuum, which can be decomposed into single-mode squeezed states of the symmetric and antisymmetric modes $\ket{\text{S}}=(\ket{1,+1} + \ket{1,-1})/\sqrt{2}$ and $\ket{\text{AS}}=(\ket{1,+1} - \ket{1,-1})/\sqrt{2}$, respectively~\cite{Kruse2016}.
For this purpose, the state $\ket{1,0}$ is first depleted by a transfer to $\ket{\text{IP}1}$ and then coherently populated by the symmetric state via a circularly polarized radio-frequency (RF) $\pi$ pulse~\cite{Kunkel2019}.
This yields a one-mode squeezed vacuum state in $\ket{1,0}$ which has a well-defined phase relation with respect to $\ket{\text{IP}1}$ and serves as the second interferometric input state $\ket{\text{IP}2}$.

The mode $\ket{\text{AS}}$ is not addressed by the RF pulse, hence does not contribute to the interferometric sequence, and is therefore disregarded.
The orientation of the squeezing can be optimally adjusted in two ways.
Either a suitable holding time imprints a phase between $\ket{1,0}$ and $\ket{2,0}$~\cite{Peise2015a}, optimally rotating the squeezing ellipse at the north pole around the $J_z$ axis, or the phases of the interferometer pulses are appropriately chosen relative to the squeezing orientation.
The optimum can be experimentally identified by measuring the output variance as a function of the squeezing orientation.

A differential phase measurement relies on two identical interferometers. First, two BECs are optically trapped at a vertical separation of $L$. Each of them is prepared in the same squeezed state at the north pole of the Bloch sphere.
The squeezed states are then released from the traps and manipulated by joint Raman light pulses.
We propose to set up each interferometer as shown in Fig.~\ref{fig4}.
Before coupling internal to momentum modes, the squeezing ellipses are rotated onto the equator (cf. Fig.~\ref{fig1}). A similar spin-space beamsplitter precedes the readout.
These transformations can be implemented by a co-propagating $\pi/2$ Raman pulse.
Similar to the Mach-Zehnder-like interferometer in Fig.~\ref{fig2}b, counter-propagating Raman $\pi$ pulses control the motional degree of freedom. The pulses simultaneously exchange $\ket{\text{IP}1}$ with $\ket{\text{IP}2}$ and kick the two states in opposite directions by transferring even multiples of the photon momentum. Thereafter, suitable Raman pulses redirect and finally stop the relative motion of the two wave packets.

In this way, a differential interferometer sensitive to inertial effects such as the Earth's gravity gradient ~\cite{McGuirk2000, Jaffe2018} or GWs can be generated by an appropriate sequence of light pulses, e.g., a six-pulse sequence as depicted in Fig.~\ref{fig4}.
Here, the interferometer employs the optimal-orientation squeezed state as described above.
The wave packets diverge for a time $T$ until a counter-propagating Raman $\pi$ pulse stops the relative motion.
Another counter-propagating $\pi$ pulse, but with inverted wave vectors, accelerates the wave packets towards each other.
As soon as spatial overlap is re-established, the relative motion of the wave packets is terminated and the full ensemble moves along a joint trajectory again.
A final co-propagating $\pi/2$ pulse couples both atomic wave packages and maps the phase onto a population imbalance which can be extracted from a spin-state resolving number measurement of the output. 
In principle, each of the two interferometers provides a squeezing-enhanced signal if the noise background is small enough.
As we outlined in Section~\ref{sec:quantum limits}, this will be especially the case for a differential analysis of both signals.

For $T'=0$ in Fig.~\ref{fig4}, the topology resembles a conventional three-pulse Mach-Zehnder interferometer as shown in Fig.~\ref{fig2}b. However, in contrast to the latter, our scheme inverts the atomic momentum by the composition of two counter-propagating Raman pulses. Therefore, the two interferometric paths correspond to different internal states and, thus, different interaction strengths. Our results in Sections~\ref{sec:optimal_squeezing} and~\ref{sec:DQL} assume that the associated phases along the two paths are, on average, equal. This can be achieved by choosing $T'=2T$.
Alternatively to reversing the wave vectors of the second central pulse, an additional co-propagating $\pi$ pulse could cancel the first exchange of internal states.
Similarly, the transfer of large momenta can be established by a sequence of Raman  $\pi$ pulses or by additional Bragg sequences~\cite{Mueller2009a,McDonald2014EPL,McGuirk2000,Plotkin2018PRL,Berg2015,Chiow2011,Gebbe2021NatComm,beguin_atom_2023,kirsten-siems_large-momentum-transfer_2023}.
Therefore, our analysis of the phase shifts occurring in differential light-pulse interferometry (Fig.~\ref{fig2}c) can be directly applied to the presented scheme. Today, our concept indeed appears to be the most straight-forward way to exploit OOS in atomic interferometry.

\section{Conclusion}
\label{sec:conclusion}

The precision of conventional, i.\,e., not quantum-enhanced atom interferometers is fundamentally limited by the SQL at $\Delta \phi_{\mathrm{SQL}}=1/\sqrt{N}$. This suggests that the sensitivity can always be improved by increasing the atom number $N$, which is technically challenging but, in principle, possible.

However, when increasing $N$, one has to decide whether to do so at constant probe density or volume (or anything in between). Large probe states require high levels of gravity-gradient compensation to avoid dephasing, while dense probes suffer from atomic interactions. For a given probe volume, the competition of shot noise and density effect yields an optimal atom number $N$ and a corresponding minimal phase uncertainty. This defines the DQL, which we introduce to facilitate the evaluation of realistic sensitivity limits. Once the probe volume is chosen such that it optimally balances different noise terms assuming the best achievable gravity-gradient compensation, the ultimate sensitivity is set by the volume-specific DQL. Using more than the corresponding optimal number of atoms impairs the sensitivity.

Quantum-enhanced atom interferometers use entanglement to improve the sensitivity without increasing $N$. However, we find that the DQL is limiting for many entangled probes such as phase-squeezed, number-squeezed, and twin-Fock states, too. Even so, we devise a method to overcome the DQL: OOS anticipates the density effect and thus allows to make full use of the initial squeezing. This is reminiscent of photonic GW detectors, where tailored squeezing optimally balances shot noise and radiation pressure. In addition to introducing the DQL and suggesting OOS as a way to surpass it, we discuss how to generate and employ optimal-orientation squeezed states with present-day experimental techniques.

Generally, our analysis of the density effect and the proposed circumvention by OOS can help to design new and extend the achievable sensitivity of existent light-pulse atom interferometers. This offers fascinating perspectives for high-precision inertial sensing, e.\,g., of the gravitational constant, for inertial navigation or for space-borne quantum gravimetry. However, most importantly, we show that the DQL will limit near-term differential atom interferometers for gravity gradiometry and GW detection. Hence, OOS is vital for advancing these fields. Due to its practical relevance, our proposal calls for an experimental proof of concept in the near future. 

\begin{acknowledgments}
  We thank K. Hammerer for a review of the manuscript and helpful discussions.
  P.F. thanks Dmytro Bondarenko for helpful discussions.
  We acknowledge financial support from the Centre for Quantum Engineering and Space-Time Research (QUEST) and the Deutsche Forschungsgemeinschaft (DFG) through CRC 1227 (DQ-mat), projects A02, B01, B07, B09.
  F.A. acknowledges support from the Hannover School for Nanotechnology (HSN).
  P.F. acknowledges funding from the Canada First Research Excellence Fund, Quantum Materials and Future Technologies Program.
  D.S. gratefully acknowledges funding by the Federal Ministry of Education and Research (BMBF) through the funding program Photonics Research Germany under contract number 13N14875.
\end{acknowledgments}

\appendix
\section{Estimating the DQL}
\label{sec:app}
We assume that the probe state is concentrated at the north pole of the (appropriately rotated) Bloch sphere. This gives rise to the one-mode approximation: we eliminate mode $a$ by expanding the Holstein-Primakoff representation~\cite{HolsteinPrimakoff1940} up to first order in mode $b$ and obtain
\begin{equation}
    J_x= \frac{\sqrt{N}}{2}(b+b^\dagger),\quad J_z= \frac{N}{2}.
\end{equation}
We further consider squeezed Gaussian probe states 
\begin{equation}
|\psi\rangle=\operatorname{e}^{\frac{1}{2}(\zeta^*b^2-\zeta b^{\dagger 2})}|0\rangle
\end{equation}
with $\zeta=s\operatorname{e}^{i\theta}$ and $s\geq 0$. The sensitivity from Eq.~\eqref{eq:sensitivity} then evaluates to
\begin{multline}
    \label{eq:sensitivityApp}
    \Delta^2\phi= \frac{1}{N}\left\{\left(1+\frac{\mu^2N^2}{4}\right)\cosh 2s\right.\\\left.-\left[\left(1-\frac{\mu^2 N^2}{4}\right)\cos\theta+\mu N\sin\theta\right]\sinh 2s\right\},
\end{multline}
where
\begin{equation}
    \mu=\frac{2U}{\hbar}\int_0^{2T}\!\operatorname{d}\!t\,\frac{1}{V(t)},
\end{equation}
$U=4\pi\hbar^2a_s/m$ and $V$ is the probe volume.

To obtain an analytical expression for $\mu$, we approximate $V(t)$ in the following way. We assume that, spatially, the probe is initially in the ground state of a spherically harmonic trap and then freely expands once it is released from the trap at time $t=0$. In particular, we model the expansion neglecting the coupling between internal and external degrees of freedom (single-spatial-mode approximation) as well as the interatomic interactions. This yields Gaussian spatial wave functions with a standard deviation $w$ that solves
\begin{equation}
    \ddot{w} =\frac{\hbar^2}{4m^2w^3},\;\;w(0)=\sqrt{2}\Delta r,\;\;\dot{w}(0)=\sqrt{2}\Delta v,
\end{equation}
cf. Ref.~\cite{PerezGarcia1996} and Appendix~\ref{sec:gaussvariation}, such that
\begin{equation}
    w(t) = \frac{1}{2\sqrt{2}\bar{m}\Delta r}\sqrt{16\bar{m}^2(\Delta r)^4+32\bar{m}(\Delta r)^2ct+(1+16c^2)t^2},
\end{equation}
where we have introduced $\bar{m}=m/\hbar$ and the dimensionless $c=\bar{m}\Delta r\Delta v$.
We identify the probe volume with a $2$-$\Delta r$-sphere, $V(t) = \frac{4}{3}\pi [\sqrt{2}w(t)]^3$, and obtain
\begin{align}
\begin{split}
    \label{eq:mu}
    \mu &= \frac{3U\bar{m}}{4\pi \hbar \Delta r}\left(\frac{4c+y}{\sqrt{1+y^2}}-4c\right),\\
    y &= \frac{T}{4cT + 2\bar{m}(\Delta r)^2}.
\end{split}
\end{align}

We can estimate the DQL by minimizing the phase variance in Eq.~\eqref{eq:sensitivityApp} over $N$. Conveniently, our approximations have made $V$ and, hence, $\mu$ independent of $N$. The result of the optimization does not depend on the phase- or number-squeezed state under consideration:
\begin{equation}
    \Delta^2\phi_{\mathrm{DQL}}=\min_N\Delta^2\phi|_{\theta\in\{0,\pi\}}=\mu
\end{equation}
with $\mu$ as in Eq.~\eqref{eq:mu}. For a coherent probe state, the DQL is attained at $N_{\mathrm{DQL}}=2/\mu$.

In our examples, see Fig.~\ref{fig2}c, $y$ is of the order of $10^{-2}$ or smaller. Taylor expanding $\mu$ around $y=0$ provides the approximation in Eq.~\eqref{eq:DQL},
\begin{equation}
    \label{eq:DQLapprox}
    \Delta^2\phi_{\mathrm{DQL}}=\frac{3U\bar{m}y}{4\pi\hbar\Delta r }(1-2cy)+\mathcal{O}(y^3).
\end{equation}
When simulating the gradiometer described in Fig.~\ref{fig2}c, we take into account that the $2$-$\Delta r$-sphere does not contain all of the $N$ atoms by using an effective $\mu'=\si{0,954}^2\mu$. Accordingly, for Fig.~\ref{fig3}, we correct the DQL from Eq.~\eqref{eq:DQL} by the same factor of $\si{0,954}^2$.

\section{Gaussian variational model for the expansion of BECs}
\label{sec:gaussvariation}

We model the spatial degrees of freedom of the wave packets traversing the atom interferometer following Ref.~\cite{PerezGarcia1996}. There, the evolution of a single-mode BEC is approximated by minimizing the energy given by the Gross-Pitaevskii equation~\cite{Pitaevskii2003} over Gaussian wave functions. However, we believe that the Lagrangian density in Ref.~\cite{PerezGarcia1996} is incorrect. Furthermore, the authors of Ref.~\cite{PerezGarcia1996} neglect---for obvious reasons---the evolution of the global phase, which is though crucial for determining the density effect. Therefore, in the following we repeat the variational analysis from Ref.~\cite{PerezGarcia1996} for spherically symmetric Gaussian wave packets evolving in free space. We formulate our ansatz in center-of-mass coordinates and include the evolution of a global phase.

The Lagrangian density corresponding to the Gross-Pitaevskii equation in free space is
\begin{equation}
    \label{eq:lagrangian}
    \mathcal{L}=i\hbar (\psi \partial_t \psi^*-\psi^*\partial_t\psi) + \frac{\hbar^2}{2m}|\nabla\psi|^2+\frac{2\pi \hbar^2a}{m}|\psi|^4.
\end{equation}
We minimize the action over wave functions of the form
\begin{equation}
    \label{eq:gaussansatz}
    \psi(x,y,z,t)=u\operatorname{e}^{i\gamma}\hspace{-1em}\prod_{\eta\in\{x,y,z\}}\hspace{-1em}\operatorname{e}^{-\frac{\eta^2}{2w^2(t)}+i\alpha(t)\eta+i\beta(t)\eta^2}
\end{equation}
with an initial normalization of $\int\!\operatorname{d}^3\!r\,|\psi(x,y,z,0)|^2=N$. To this end, we plug Eq.~\eqref{eq:gaussansatz} into Eq.~\eqref{eq:lagrangian} and evaluate the Lagrangian $L=\int\!\operatorname{d}^3\!r\mathcal{L}$. The Euler-Lagrange equations $\frac{\partial L}{\partial q}=\frac{\operatorname{d}}{\operatorname{d}\!t}\frac{\partial L}{\partial (\partial_t q)}$ for the real parameters $q\in\{u,\alpha,\beta,\gamma,w\}$ then yield
\begin{align}
    u^2 &= \frac{N}{\sqrt{\pi}^3w^3}\\
    \alpha &= 0\\
    \beta &= \frac{m\partial_t w}{\hbar w}\\
    \partial_t\gamma &= -\frac{7\hbar a N}{4\sqrt{2\pi}m}\frac{1}{w^3}-\frac{3\hbar}{4m}\frac{1}{w^2}\\
    \partial_t^2w &= \frac{\hbar^2}{4m^2}\frac{1}{w^3}+\frac{\hbar^2aN}{2\sqrt{2\pi}m^2}\frac{1}{w^4}.
\end{align}

\bibliography{main.bib}
\end{document}